\documentclass[aps, superscriptaddress, groupedaddress, floatfix, reprint]{revtex4-2} 
\usepackage{graphicx}
\usepackage{dcolumn}
\usepackage{bm}
\usepackage{amsfonts}
\usepackage{amsmath}
\usepackage{amssymb}
\usepackage{color}
\usepackage[normalem]{ulem}
\usepackage[percent]{overpic}
\usepackage{subcaption}
\usepackage{tikz}
\usepackage{graphicx}

\usepackage[colorlinks=true,citecolor=blue]{hyperref}
\hypersetup{colorlinks=true,citecolor=blue,linkcolor=red,urlcolor=blue}

\def\be{\begin{equation}}
\def\ee{\end{equation}}

\def\rv{{\bf r}}
\def\kv{{\bf k}}

\def\vv{{\bf v}}
\def\Ev{{\bf E}}
\def\Bv{{\bf B}}
\def\Fv{{\bf F}}
\def\Jv{{\bf J}}
\def\nablav{{ \nabla}}

\def\Lambdav{{\bm \Lambda}}
\def\lambdav{{\bm \lambda}}

\begin{document}
\title{Magneto-thermoelectricty of anisotropic two-dimensional materials}
\author{Maryam Nezafat}
\affiliation{Department of Physics, Institute for Advanced Studies in Basic Sciences (IASBS), Zanjan 45137-66731, Iran}
\author{Shahin Barati}
\affiliation{Department of Physics, Institute for Advanced Studies in Basic Sciences (IASBS), Zanjan 45137-66731, Iran}
\author{Saeed H. Abedinpour}
\email{abedinpour@iasbs.ac.ir}
\affiliation{Department of Physics, Institute for Advanced Studies in Basic Sciences (IASBS), Zanjan 45137-66731, Iran}

\date{\today}

\begin{abstract}
We use semiclassical Boltzmann transport theory to analytically study the electronic contribution to the linear thermoelectric response of anisotropic two-dimensional materials subjected to a perpendicular magnetic field.
Conventional methods, such as the relaxation-time approximation within the Boltzmann formalism, often yield qualitatively incorrect results when applied to anisotropic mediums. 
On the other hand, the vector mean free path provides exact solutions for the linearized Boltzmann transport equation, and in principle, we can systematically evaluate it to any desired order in the external magnetic field. 
After laying out the general formalism of the magneto-thermoelectric responses of anisotropic systems based on the vector mean free path approach, we study the longitudinal and Hall charge conductivities, as well as the Seebeck and Nernst coefficients of two representative anisotropic two-dimensional electronic systems: a semi-Dirac semimetal and a two-dimensional system of tilted massless Dirac fermions.
\end{abstract}
\maketitle

\section{Introduction}\label{introduction}

The ongoing miniaturization of electronic devices, the surge in energy demands, and severe environmental crises, such as global warming, all call for increased energy efficiency. 
As a potential solution to these problems, thermoelectricity, which investigates the interconversion of heat and charge flow, is gaining rising attention. Despite being a centuries-old science with vast theoretical and practical breakthroughs, when compared with purely electronic phenomena, thermoelectricity is neither theoretically well understood nor are its full practical capacities utilized~\cite{behnia_book}. In particular, the quest for high-quality thermoelectric materials has been mainly in vain. From the theoretical point of view, one big challenge is the competition between different phenomena that contribute to the overall performance of a thermoelectric device. 

Thermoelectricity generally studies the response of a material to electric fields and temperature gradients. Electric, thermoelectric, and thermal conductivities relate these external perturbations to the electric and heat currents. When a magnetic field is also present, Hall and Nernst's effects appear as the additional transverse responses in the system~\cite{behnia_book,behnia_jpcm2009,behnia_rpp2016}. 

Following the momentous isolation of a single layer graphene in 2004~\cite{novoselov_science2004,neto2009electronic}, a vast family of two-dimensional (2D) materials identified~\cite{Miro_CSR2014}. These atomically thin materials span a wide range of different physical properties. Identifying a family of structures with desired electronic properties, from metals and semimetals to semiconductors and insulators, is relatively easy. Both non-magnetic and magnetic materials and the ones with strong spin-orbit coupling or topologically non-trivial properties have their place on this counter~\cite{Khazaei_JMCC2017,Das_ARMR2015, Burch_nature2018,Bernevig_Nature2022,Sangwan_ARCP2018,Avsar_RMP2020}. 

Among different 2D systems, Dirac fermions have a particular place due to their extraordinary transport properties~\cite{neto2009electronic, kotov2012electron, wehling2014dirac,qi2011topological,heckelsky2009thermopower}.
Dirac materials are systems where low-energy electrons obey the relativistic Dirac equation~\cite{vafek2014dirac,armitage2018weyl}.
Major discovered classes of two and three-dimensional Dirac materials include topological insulators~\cite{bernevig2006quantum,wan2011topological} as well as Weyl~\cite{burkov2011weyl,xu2016optical} and Dirac semimetals~\cite{young2012dirac,wang2013three}, whose conduction and valence bands touch each other at isolated points in the Brillouin zone. Such band crossings are topologically protected, i.e., cannot be removed by perturbations on the Hamiltonian without breaking its preserving symmetries~\cite{yang2014classification, yang2018symmetry, gao2019topological}.
Topological semimetals' transport and optical properties in the presence of electric and magnetic fields have been studied theoretically and experimentally~\cite{ashby2014chiral, tabert2016optical, barati2017optical}. 
More recently, the thermoelectric properties of these materials have gained increasing interest too~\cite{zhu2010universal,lundgren2014thermoelectric,chen2016thermoelectric,liang2017anomalous,barati2020thermoelectric}.

Another subject of great interest in two-dimensional Dirac systems is the quest for materials in which the Dirac cone is deformed. 
Several properties of materials with tilted Dirac cones~\cite{Moradpouri_JHEP2023,Faraei_PRB2020,goerbig2008tilted,yang2018effects,redondo2021quantum,cheng_pccp2017} and semi-Dirac systems where electrons display linear or quadratic dispersion depending on their direction of propagation~\cite{dietl2008new,banerjee2009tight, pardo2010metal,adroguer2016diffusion,real2020semi,banerjee2012phenomenology}  have been extensively studied. All such systems are anisotropic; therefore, one naturally expects anisotropy in their different responses.

 Anisotropy, either intrinsic (due to the electronic structure) or extrinsic (due to anisotropic scatterings), generally adds an extra dimension to the pool of the transport properties~\cite{sabzalipour_JPCM2015, Trushin_PRB2019, barati2020thermoelectric}. 
However, appropriate treatment of such an anisotropy requires extra attention even in studying simple charge transport in the absence of magnetic field~\cite{Trushin_PRB2019,kim2019vertex,Vyborny_PRB2009,sabzalipour_JPCM2015}. 

The current work investigates the magneto-thermoelectric responses of anisotropic two-dimensional materials within the semiclassical Boltzmann formalism. We aim to find an exact solution for the Boltzmann transport equation (BTE) that is linearized in the external electric field and temperature gradient. Subjecting an anisotropic material to an external magnetic field further complicates the problem.
We introduce an Ansatz for the non-equilibrium distribution function in terms of the vector mean free path (VMFP)~\cite{pikulin_PRB2011}. 
We expand VMFP in the powers of the magnetic field and find a set of closed equations that can, in principle, be solved recursively to the desired order in the magnetic field.

We assess our proposed approach by applying it to two generic anisotropic model systems, i.e., a 2D semi-Dirac semimetal and a 2D system of tilted massless Dirac fermions. We consider scattering from short-range isotropic impurities and find these two models' vector mean free paths. Then, we calculate the longitudinal and transverse magneto-thermoelectric responses to leading order in the magnetic field as functions of the Fermi energy. 

The rest of this paper is organized as follows. 
In Sec.~\ref{sec:boltzmann}, we introduce the semiclassical Boltzmann transport formalism for anisotropic systems and discuss how we solve it utilizing VMFP in the presence of a magnetic field.
We explain how we can find electric, thermal, and thermoelectric conductivities from the vector mean free paths in Sec.~\ref{sec:thermoelectric response}. 
Then, in Sec.~\ref{sec:examples}, we introduce two model systems that we apply our formalism to study their magneto-thermoelectric responses.
We summarize and conclude our work in Sec.~\ref{sec:conclusion}. 
Three appendices at the end provide details of the calculations that lead to the results presented in the main body of the paper.

\section{Boltzmann transport equation for anisotropic materials}\label{sec:boltzmann}
This section outlines the semiclassical Boltzmann transport equation for an \emph{anisotropic} system subjected to external electric field $\Ev$, magnetic field $\Bv$, and temperature gradient $\nablav T$.
Introducing vector mean free paths, we find an exact solution of the BTE to linear order in the electric field and temperature gradient and for an arbitrary strength of the magnetic field.

We start with the BTE for an electronic wave packet subjected to uniform external electric
and magnetic fields and temperature gradients~\cite{Girvin_book}
\be \label{eq:BTE}
\frac{\partial f}{\partial t} +\dot{\rv} \cdot \nablav_{\rv}f+ \dot{\kv} \cdot \nablav_{\kv}f = \left(\frac{\partial f}{\partial t} \right)_{\rm coll.} ,
\ee
where $f(\kv, \rv)$ refers to the local distribution of electrons with wave
vector $\kv$ in the neighborhood of position $\rv$, and $\left(\frac{\partial f}{\partial t} \right)_{\rm coll.}$ represents the change in the distribution function due to scatterings.
The semiclassical equations of motion, in the absence of anomalous velocity~\cite{Girvin_book}, are
\be \label{eq:mov}
\begin{aligned}
\dot{\rv}&=\vv_\kv=\dfrac{1}{\hbar}\nablav_{\kv} \varepsilon_\kv,\\
\dot{\kv}&=-\dfrac{e}{\hbar}\left(\Ev+\vv_{\kv}\times \Bv \right)  .
\end{aligned}
\ee
For elastic scatterings, with the help of Fermi's golden rule, the scattering term reads
\be \label{eq:coll}
\left(\frac{\partial f}{\partial t} \right)_{\rm coll.}=\sum_{\kv'}  w_{\kv , \kv'}\left[f(\kv')-f(\kv)\right],
\ee 
where the scatterings rate, within the first Born approximation, is
 \be
w_{\kv ,\kv'}=\dfrac{2\pi}{\hbar}n_{\rm imp}|V_{\kv\kv'}|^2\delta(\varepsilon_{\kv}-\varepsilon_{\kv'}).
\ee
Here, $n_{\rm imp}$ is the density of impurities in the sample, and $V_{\kv\kv'}$ is the impurity potential describing the scattering of an electron from state $\kv$ to $\kv'$. 

In the steady state $\partial f/\partial t= 0$, if we define the deviation from equilibrium distribution as $g_{\kv} \equiv f_{\kv} - f^0_{\kv} $, where $f^0(\varepsilon)=1/[e^{(\varepsilon-\mu)/(k_{\rm B}T)}+1]$ is the equilibrium Fermi-Dirac distribution function (with $\mu$ the chemical potential), to leading order in the electric field and temperature gradient, we find
\be
\nablav_{\rv} f \approx \nablav_{\rv} f^0=\left(-\frac{\partial f^0}{\partial {\varepsilon}}\right) 
\left(\dfrac{\varepsilon -\mu}{T}\right){\nablav} T,
\ee
and
\be
\nablav_{\kv} f=\nablav_{\kv} f^0+\nablav_{\kv} g=
\left(\frac{\partial f^0}{\partial {\varepsilon}}\right)  \nablav_{\kv} \varepsilon_\kv+\nablav_{\kv} g_\kv.
\ee
Now, introducing
\be
\Fv = \left(-\frac{\partial f^0}{\partial {\varepsilon}}\right)
\left( e \Ev +\dfrac{\varepsilon -\mu}{T}\,\nablav T \right) ,
\ee
as the generalized perturbation, the Boltzmann transport equation~\eqref{eq:BTE}, in the steady state limit and to the leading order in the perturbation $\Fv$, becomes
\be \label{eq:BTE_lin}
\vv_{\kv}\cdot {\bf F} =\sum_{\kv'} w_{\kv , \kv'}\left(g_{\kv'}-g_\kv\right)- \widehat \Omega_\kv g_\kv ,
\ee
where $\widehat \Omega_\kv = -e(\vv_\kv\times \textbf{B}) \cdot \nablav_{\kv}/\hbar$, and we have used $\varepsilon_\kv=\varepsilon_{\kv'}$, and therefore $f^0(\kv) = f^0(\kv')$, for elastic scatterings.

Now, for the non-equilibrium part of the distribution function, to linear order in the perturbation $\Fv$, we take the following general Ansatz
\be \label{eq:gk}
g_\kv= - {\bf F} \cdot \Lambdav_{\kv} ,
\ee
in terms of the vector mean free path $\Lambdav_\kv$~\cite{pikulin_PRB2011}, which depends on the magnetic field $\Bv$, and in practice we can determine it solving the linearized BTE~\eqref{eq:BTE_lin}.
The linearized BTE in terms of the VMFP reads
\be \label{eq:lamB}
\left({\overline w}_\kv+\widehat\Omega_\kv\right)\Lambdav_{\kv}=\vv_\kv+\sum_{\kv'}w_{\kv,\kv'}\Lambdav_{\kv'},
\ee
which is an integrodifferential equation in terms of $\Lambdav_{\kv}$, and
${\overline w}_\kv=\sum_{\kv'}w_{\kv,\kv'}$ is the scattering rate averaged over all outgoing states.
Note that in the absence of an external magnetic field, we have $\widehat\Omega=0$. Then, it is customary to define direction-dependent relaxation times as $\tau^i_\kv=\Lambda^i_\kv/v^i_\kv$, with $i=x,y,z$, and find the familiar expression~\cite{kim2019vertex,sabzalipour_JPCM2015, barati2020thermoelectric,Vyborny_PRB2009} 
\be \label{eq:relaxtimeaniso1}
{\overline w}_{\kv} \tau^i_\kv=1+\frac{1}{v^i_{\kv}}\sum_{\kv'}w_{\kv,\kv'}v^{i}_{\kv'}\tau^{i}_{\kv'}.
\ee

\subsection{Two-dimensional medium subjected to a perpendicular magnetic field}\label{subsec:TDB}
We are interested in the thermo-magnetic properties of two-dimensional materials subjected to a perpendicular uniform magnetic field. Taking the 2D material in the $x-y$ plane and considering $\Bv=B \hat z$, it is easy to show
$\widehat\Omega_\kv=(eB/\hbar)\hat{z}\cdot\left(\vv_\kv\times \nablav_\kv \right)$.
If the magnetic field is not too strong, we can expand VMFP in powers of the magnetic field as
\be \label{eq:La}
\Lambdav_{\kv}=\sum_{s=0}^\infty \lambdav^{(s)}_{\kv}{B}^s ,
\ee
and the problem reduces to finding $\lambdav^{(s)}_{\kv}$, which are intrinsic properties of the system independent of the external perturbations.

Upon the substitution of  Eq. (\ref{eq:La}) into Eq. (\ref{eq:lamB}) and equating the coefficients of the corresponding powers of the magnetic field on both sides of the expression, we find
\be \label{eq:lams}
\begin{split}
\overline{w}_{\kv}\lambdav_{\kv}^{(0)}&=\vv_\kv+\sum_{k'}w_{\kv,\kv'}\lambdav_{\kv'}^{(0)},\\
\overline{w}_{\kv}\lambdav_{\kv}^{(s)}&=\sum_{\kv'}w_{\kv,\kv'}\lambdav_{\kv'}^{(s)}-\frac{e}{\hbar}\hat{z} \cdot (\vv_\kv \times \nablav_\kv)\lambdav_{\kv}^{(s-1)},~~ s\geq 1 .
\end{split}
\ee
In principle, we can solve this sequence of equations to the desired order in $s$. At each order, all parameters, except the coefficient $\lambdav_{\kv}^{(s)}$ itself, are known from the solution of equations for smaller powers.

Furthermore, note that the particle number conservation requires $\sum_\kv g_\kv=0$. As this should hold irrespective of  perturbation direction and magnetic field strength, we should demand 
 \be \label{eq:number}
\int_{\varepsilon}\mathrm{d}\Omega \lambda^{i,(s)}_{\kv}=0, \quad \forall \quad i,s ,
\ee
where the integration is over the constant energy surface $\varepsilon$.
 
\section{Thermo-magnetic linear response coefficients}\label{sec:thermoelectric response}
Once we have the non-equilibrium distribution function from the solution of BTE at our disposal, we can deduce different linear transport coefficients from the general relations\be\label{eq:JeJq}
\begin{split}
\Jv_e&=-e\sum_\kv f_k \vv_\kv,\\
\Jv_q&=\sum_\kv f_k \left(\varepsilon_\kv-\mu\right)\vv_\kv,
\end{split}
\ee
for the charge $\Jv_e$ and heat $\Jv_q$ current densities.
Thermo-magnetic linear responses are defined as the coefficients of the relation between current densities and the external perturbations~\cite{Girvin_book,behnia_book,behnia_jpcm2009}
\be\label{eq:current}
\left(
\begin{array}{c} 
\Jv_e\\
\Jv_q\\
\end{array} \right) =
\left(\begin{array}{c c} 
 {\sigma} & {\alpha}\\
T{\alpha} &{\kappa}\\
\end{array} \right) \left(
\begin{array}{c} 
{\bf E}\\
-{ \nabla} T\\
\end{array} \right),
\ee
where ${\sigma}$, ${\alpha}$, and ${\kappa}$ are tensors of the electrical, thermoelectric, and heat conductivities, respectively.
These responses are conveniently defined in terms of the kinetic coefficients, $\sigma\equiv\mathcal{L}^0$, $\alpha\equiv-\mathcal{L}^1/(eT)$, and $\kappa\equiv\mathcal{L}^2/(e^2T)$~\cite{Girvin_book,behnia_book}, with
\be \label{eq:transport1}
\mathcal{L}^n=\int \mathrm{d}\varepsilon \left(-\dfrac{\partial f^0}{\partial \varepsilon}\right)\Sigma(\varepsilon)(\varepsilon-\mu)^n,\quad n=0,1,2,
\ee
where
\be\label{eq:general_transport}
\Sigma_{ij}(\varepsilon)=\frac{e^2}{A}\sum_\kv \delta(\varepsilon-\varepsilon_\kv)v^i_\kv  \Lambda^j_\kv,
\ee
is the generalized transport distribution function~\cite{mahan_pnas1996}.

At low temperatures $k_{\rm B}T\ll \mu$,  assuming that $\Sigma(\varepsilon)$ is a regular and differentiable function of energy, we can evaluate the integrals of Eq.~\eqref{eq:transport1} with the help of the Sommerfeld expansion, and to the leading-orders in temperature we find
\be \label{eq:lowT}
\begin{aligned}
&\sigma
\approx \Sigma(\mu),\\
&\alpha
\approx -\frac{\pi^2}{3e}k^2_{\rm B}T\, \Sigma'(\mu),\\
& \kappa
\approx\frac{\pi^2}{3e^2}k^2_{\rm B}T\, \Sigma(\mu),
\end{aligned}
\ee
where $\Sigma'(\mu)= \mathrm{d}\Sigma(\mu)/\mathrm{d}\mu$. Note that the Mott relation and the Wiedemann-Franz law are satisfied for both longitudinal and transverse components of the thermoelectric responses within the set of approximations that led to Eq.~\eqref{eq:lowT}.

If we set the charge current equal to zero, Eq.~\eqref{eq:JeJq} gives the induced electric field in response to an applied temperature gradient as $\Ev= S \nabla T$, with the generalized Seebeck tensor defined as $S=\sigma^{-1}\cdot \alpha$. The diagonal elements of the Seebeck tensor are the usual thermopower, giving the electric field induced parallel to the temperature gradient. In contrast, its off-diagonal elements are the Nernst coefficients~\cite{behnia_book, behnia_jpcm2009}, and give the electric field induced perpendicular to the applied temperature gradient. 
Also, notice that the electronic contribution to thermal conductivity, measured in the vanishing charge current limit, is 
$\kappa_e=\kappa-T\alpha \cdot S\approx \kappa +{\cal O}(T^3)$.

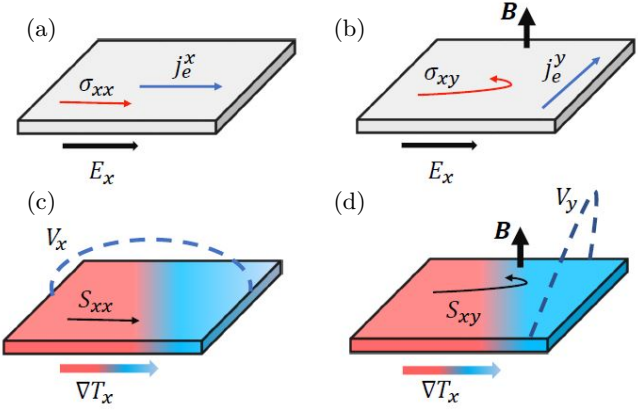
\begin{figure}
\centering
\begin{tikzpicture}
\node at (0,0)
{\includegraphics[width=\linewidth]{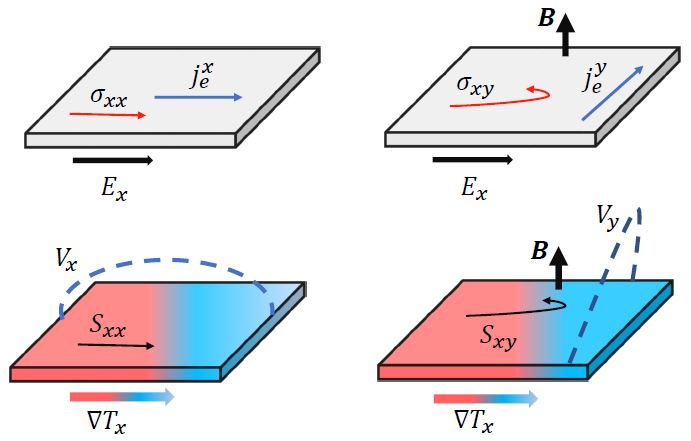}};
\draw (-3.7,2.3) node [font=\small] {(a)};
\draw (0.4,2.3) node [font=\small] {(b)};
\draw (-3.7,0.) node [font=\small] {(c)};
\draw (0.4,0) node [font=\small] {(d)};
\end{tikzpicture}
\caption{A schematic illustration of different magneto-thermoelectric responses. 
(a) An electric current induced parallel to the applied electric field $E$ is given by the longitudinal components of the electric conductivity $\sigma_{ii}$. (b) With a finite magnetic field $B$, the Hall response $\sigma_{xy}$ generates a charge current transverse to the applied electric field.
Seebeck $S_{ii}$ (c) and Nernst $S_{xy(yx)}$ (d) coefficients, respectively, give the longitudinal and transverse electric fields generated in a sample subjected to a temperature gradient $\nabla T$ and a perpendicular magnetic field $B$.
\label{fig:thermo_schematic}}
\end{figure}

In Fig.~\ref{fig:thermo_schematic}, we provide a schematic illustration of several magneto-thermoelectric phenomena. 

\section{Anisotropic two-dimensional semimetals}\label{sec:examples}
Having outlined our formalism for the VMFP and the thermoelectric coefficients, in this section, we turn to inspect the magneto-thermoelectric properties of two specific examples of anisotropic materials in the presence of short-range impurity scatterings. 
First, we study a 2D semi-Dirac semimetal and then look into a 2D system of tilted massless Dirac fermions. For each model, we find analytic expressions for the vector mean free path up to linear order in the external magnetic field. Then, we calculate the electrical conductivity and Seebeck tensors. We also provide a comparison of our results for the response functions with the predictions of the widely used constant relaxation time approximation (CRTA).

\subsection{Two-dimensional semi-Dirac semimetal}\label{sec:responsesemi}
We consider an effective low-energy Hamiltonian for the two-dimensional semi-Dirac semimetal, with quadratic and linear band dispersions along the x and y directions, respectively~\cite{banerjee2012phenomenology}
\be\label{eq:hamil1}
{\cal H}=\dfrac{\hbar^2 k^2_x}{2m}\hat{\sigma}_{x}+\hbar v k_y\hat{\sigma}_{y},
\ee
where $m$ is the effective band mass along the $x$ direction, $v$ is the Fermi velocity in the $y$ direction, and $\sigma_i$ (with $i=x,y$) are the Pauli matrices, acting on a pseudo-spin degree of freedom.
The energy spectrum of the Hamiltonian \eqref{eq:hamil1} is readily obtained as  
\be \label{eq:energy}
\varepsilon_{\kv ,s} =s\frac{\hbar^2}{2m}\sqrt{k^4_{x}+\gamma^2 k^2_y},
\ee
with $s=\pm$ referring to the conduction (+) and valance (-) bands, and  $\gamma = 2mv/\hbar$. 
In the following, we adopt the transformations
\be \label{eq:Polar}
 \begin{aligned}
& k_x={\rm sgn}(\cos\theta)\sqrt{2m \varepsilon\left|\cos\theta \right|}/\hbar , \\
& k_y=\varepsilon\sin\theta/(\hbar v)  ,
\end{aligned}
\ee
to find the dispersions simplified to $\varepsilon_{\kv ,s}=s \,\varepsilon$, and the corresponding eigenstates as
\be\label{eq:eigenstate}
\psi_{\kv,s}(\rv)
=\dfrac{1}{\sqrt{2A}}
\left(
\begin{array}{c} 
1\\
s e^{i\varphi_{\kv} }\\
\end{array} \right) e^{i\kv\begin{tiny}
\cdot
\end{tiny}\rv},
\ee
where $A$ is the sample area, and
$\varphi_{\kv} =\arctan\left(     {\sin\theta} / { \left|\cos\theta \right| }\right)$.
Furthermore, we find the components of the band velocity as 
\be \label{eq:velocity}
\begin{aligned}
&v^x_{\kv,s}=s \cos\theta \sqrt{2 \varepsilon \left|\cos\theta \right|/m},\\
&v^y_{\kv,s}=s v \sin\theta.
\end{aligned}
\ee
The Jacobian determinant of the transformation~\eqref{eq:Polar} is 
$\mathcal{J}(\varepsilon,\theta)=\sqrt{ m \varepsilon/(2\left|\cos\theta\right|)}/(\hbar^2  v)$, and for 
the density of states (DOS), we find
\be\label{eq:dos1}
\rho(\varepsilon)=g \sum_s \int \frac{d^2k}{(2\pi)^2}\delta(\varepsilon-\varepsilon_{\kv,s})
=\rho_0 \sqrt{\left|\varepsilon\right|/\varepsilon_0} .
\ee 
Here, $g$ is the band degeneracy factor, $\varepsilon_0 = mv^2/2$, and $\rho_0= g m K(1/2)/(\sqrt{2}\pi^2\hbar^2)$,
where $K(x)$ is the complete elliptic integral of the first kind~\cite{Gradshteyn}. Note that the energy dependence of Eq.~\eqref{eq:dos1} is similar to the DOS of a conventional three-dimensional electron gas.

\subsubsection{Vector mean free path  $\Lambda_{\kv}$}
\begin{figure}
\begin{tabular}{cc} 	
\hspace{-0.1\linewidth}\includegraphics[width=0.8\linewidth]{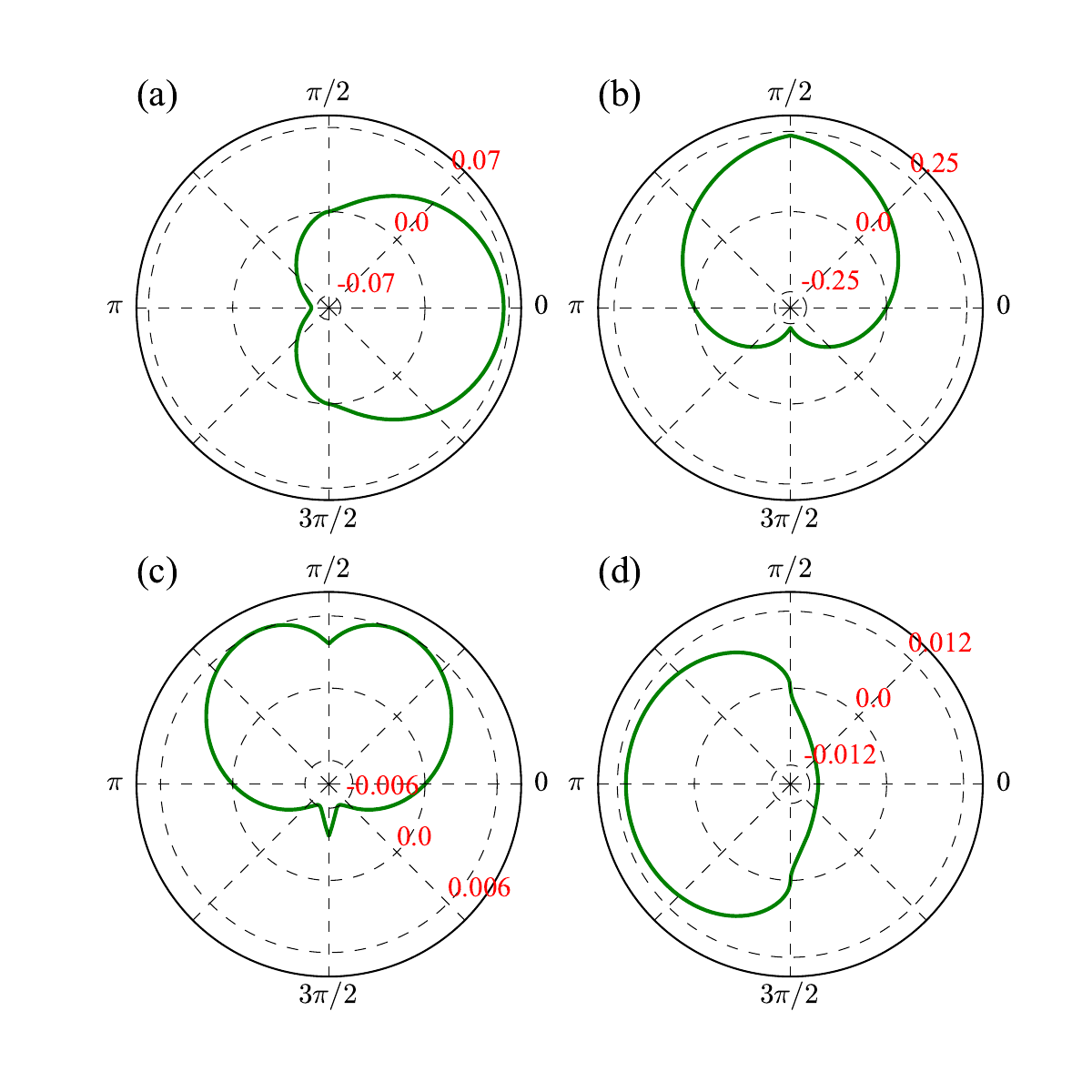}&
\hspace{-0.07\linewidth}\raisebox{0.18\linewidth}{\includegraphics[width=0.25\linewidth]{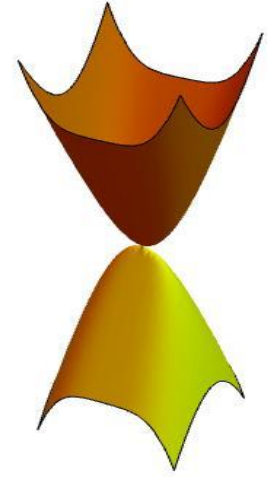}}
\end{tabular}
\caption{The polar plots of different components of the mean free path of a semi-Dirac semimetal. Top panels: zeroth order components $\lambda^{x,(0)}/l_0$ (a), and $\sqrt{\widetilde \varepsilon}\,\lambda^{y,(0)}/l_0$  (b). Bottom panels: first-order corrections
$\widetilde \varepsilon\,\lambda^{x,(1)}/(l_0 S_0)$ (c), and ${\widetilde \varepsilon^{3/2}}\lambda^{y,(1)}/(l_0 S_0)$  (d).
The sketch on the right illustrates a schematic plot of the semi-Dirac semimetal's band structure.\label{fig:lambda_semidirac}}
\end{figure}

Now we consider short-ranged impurities with random uncorrelated distribution as the source of scattering for electrons, i.e., $V(r)=V_0\sum_j\delta(\rv-{\bf R}_j)$, where ${\bf R}_j$ is the location of the $j$-th impurity. 
At weak magnetic fields, we only need to find the components of the mean free path to linear order in the magnetic field strength, i.e., $\Lambdav_\kv\approx \lambdav^{(0)}+B \lambdav^{(1)}_\kv$.
In the following, without losing the generality of our discussion, we consider electron-doped systems and drop the band index from our results. After some straightforward but cumbersome algebra (see Appendix~\ref{app:I} for the details), we find
\be \label{eq:semidirac_lambda0}
\begin{aligned}
\lambda^{x,(0)}_{\kv}&=l_0\frac{\cos\theta \sqrt{\left|\cos\theta\right|}}{I_1+I_2\left|\cos\theta \right|},\\
\lambda^{y,(0)}_{\kv}&=l_0\frac{\sin\theta}{\sqrt{\widetilde \varepsilon}\left(1-I_3\right)\left(I_1+I_2\left|\cos\theta \right|\right)},
\end{aligned}
\ee
for the components of the zeroth order mean free path, where $l_0= v\tau_0$,  with $\tau_0= (2\pi\hbar)^2/(u_0m)$ ($u_0=\pi n_{\rm imp}V^2_0/\hbar$), and $\widetilde {\varepsilon}=\varepsilon/\varepsilon_0$.
The dimensionless constants  $I_i$ are defined in Appendix~\ref{app:I}. 
Notice that our zeroth order results for the VMFP agree with the findings of Ref.~\cite{adroguer2016diffusion}.
Furthermore, the first-order correction to the mean free path, due to the magnetic field, reads
\be \label{eq:lxs}
\begin{aligned}
\lambda^{x,(1)}_{\kv}=&\frac{l_0 S_0\sin\theta}{2\widetilde \varepsilon\left(I_1+I_2\left|\cos\theta \right|\right)}\\
&\times\left[\frac{I_4}{1-I_3}+\frac{\left(3I_1+I_2\left|\cos\theta \right|\right)\left|\cos\theta\right|}{\left(I_1+I_2\left|\cos\theta \right|\right)^2}\right],\\
\lambda^{y,(1)}_{\kv}=&-
\frac{l_0 S_0\left(I_1\left|\cos\theta \right|+I_2\right)\cos\theta}{{{\widetilde \varepsilon}^{3/2}}\left(1-I_3\right)\left(I_1+I_2\left|\cos\theta \right|\right)^3\sqrt{\left|\cos\theta\right|}},
\end{aligned}
\ee
where $S_0= 2e\tau_0/m$ has the dimensions of the inverse magnetic field.
In Fig.~\ref{fig:lambda_semidirac}, we illustrate the direction dependence of the vector mean free paths, i.e., Eqs.~\eqref{eq:semidirac_lambda0} and~\eqref{eq:lxs}.
It is also easy to verify that both equations satisfy the particle number conservation, i.e.,  Eq.~\eqref{eq:number}.

\subsubsection{Thermoelectric coefficients}
Now, we present our results for the electrical and thermoelectric conductivities. 
Making use of Eq.~\eqref{eq:general_transport}, to leading order in the magnetic field we find
\be \label{sigmasd}
\sigma=\sigma_0\begin{pmatrix}
 c_1 {\widetilde \mu} &  -c_3 S_0 B/{\sqrt{\widetilde \mu}}  \\
 c_3 S_0 B /{\sqrt{\widetilde \mu}} & c_2 \\
\end{pmatrix},
\ee
for the electrical conductivity, where 
$\sigma_0= g e^2 v^2/(2u_0)$, ${\widetilde \mu}=\mu/\varepsilon_0$,
and the expressions and values of the dimensionless constants $c_i$ are given in Appendix~\ref{app:I}.
The thermoelectric conductivity tensor reads
\be \label{eq:alphasemidirac}
\alpha =-\dfrac{\pi^2 k^2_B T \sigma_0 }{6e \varepsilon_0} \begin{pmatrix}
 2c_1 &  c_3 S_0 B/{{\widetilde \mu}^{3/2}} \\
-c_3 S_0 B/{{\widetilde \mu}^{3/2}} & 0 \\
\end{pmatrix}.
\ee
Finally, for the generalized Seebeck coefficient, we obtain
\be
\label{eq:semidirac}
S=-\dfrac{\pi^2 k^2_B T}{6e \mu}
 \begin{pmatrix}
2  & c_3 {S_0B} /(c_1 {{\widetilde \mu}^{3/2}}) \\
 -3c_3 S_0B /(c_2 \sqrt{\widetilde \mu}\,) &0\\
\end{pmatrix}.
\ee
The Seebeck coefficient along the $y$-direction, where energy dispersion is linear, vanishes when we consider short-range scatterings alone.

In many ab initio calculations of the thermoelectric properties for actual materials, it is customary to use the constant relaxation time approximation~\cite{Madsen_CPC2006, Madsen_CPC2018, Pizzi_CPC2014,zebarjadi_prb2021}. 
In TABLE~\ref{table:semiDirac}, we compare the chemical potential dependence of our results for the electrical conductivity and Seebeck and Nernst responses with what one finds from the CRTA (see Appendix~\ref{app:CRTA} for the details of CRTA). The CRTA fails to provide a qualitatively accurate description of the magneto-thermoelectric effect in a semi-Dirac system.  Therefore, it seems essential to incorporate the full energy and direction dependence of the scatterings in the calculations of the thermoelectric properties. 
We should mention that there are recent efforts to go beyond the relaxation time approximation in thermo-magnetic calculations based on first principles data~\cite{rezaei_CMS2022,rezaei_CMS2023}.

\begin{table}
\caption{The chemical potential dependence of charge conductivity $\sigma_{ij}$, Seebeck $S_{ii}$, and Nernst $N_{i\neq j}$ coefficients of a two-dimensional semi-Dirac semimetal obtained from the vector mean free path (VMFP) method. The constant relaxation time approximation (CRTA) results are also provided for comparison.
\label{table:semiDirac}}
\begin{tabular}{|c|c|c|c|c|c|c|c|c|}
\hline
 Approach & $\sigma_{xx}$ & $\sigma_{yy}$  & $\sigma_{xy}(-\sigma_{yx})$   & $S_{xx}$ & $S_{yy}$  & $S_{xy}$  & $S_{yx}$\\
\hline
VMFP & $\mu$ & $\mu^0$ &$\mu^{-1/2}$ &  $\mu^{-1}$ & 0 & $\mu^{-5/2}$ & $\mu^{-3/2}$\\
\hline
CRTA & $\mu^{3/2}$ & $\mu^{1/2}$ & $\mu^{1/2}$ & $\mu^{-1}$ & $\mu^{-1}$ & 0 & $\mu^{-1}$\\
 \hline
\end{tabular}
\end{table}

\subsection{Tilted two-dimensional massless Dirac fermions}\label{sec:responsetilt}
Our second example is a two-dimensional system with anisotropic tilted Dirac cones, whose effective low-energy Hamiltonian is given by~\cite{yang2018effects}
\be\label{eq:hamiltilt}
{\cal H}=\hbar v_{x}k_{x}\hat{\sigma}_{x}+ \hbar v_{y}k_y\hat{\sigma}_{y}+\hbar \vv_{t}\cdot \kv \hat{\sigma}_{0},
\ee
where $v_{x(y)}$  is the Fermi velocity of an un-tilted but anisotropic system in the $x(y)$ direction, $\vv_{t}$  is the tilt velocity vector, 
$\sigma_{x,y}$ are the Pauli matrices,
and  $\hat{\sigma}_{0}$ is a $2\times 2$ identity matrix. 
In the following, we define our $x$ axis along the tilt direction and set $v_{t}= t v_{x}$ where $t$ is the dimensionless tilt parameter. Here, we focus only on the $0\leq  t < 1$ regime.
The spectrum of the Hamiltonian~\eqref{eq:hamiltilt} is given by 
\be \label{eq:energytilt}
\varepsilon_{\kv, s}=\hbar t v_{x}k_{x}+s \hbar \sqrt{v_x^2k_x^2+v_y^2k_y^2}.
\ee
where $s =\pm $ labels the conduction and valence bands, and the corresponding eigenstates are
\be \label{statetilt}
\psi_{\kv,s}(\rv)=\frac{1}{\sqrt{2A}}\left(
\begin{array}{ccc}
1 \\
s e^{-i\theta}\\
\end{array}
\right)  e^{i\kv
\cdot
\rv},
\ee
with $\theta =\arctan\left[{v_{y}k_{y}}/({v_{x}k_{x}})\right]$.
Note that for $t=0$ and equal Fermi velocities along $x$ and $y$ directions, the Hamiltonian~\eqref{eq:hamiltilt} reduces to the low-energy Hamiltonian of a single-layer graphene. 
Now, with the following transformations  (focusing on the electron-doped systems, i.e. $\varepsilon>0$)
\be\label{Polar2}
\begin{aligned}
k_{x}=&\frac{\varepsilon\cos\theta}{\hbar v_{x}(1+t\cos\theta)},\\
k_{y}=&\frac{\varepsilon\sin\theta}{\hbar v_{y}(1+t\cos\theta)},	
\end{aligned}
\ee
components of the band velocity read
\be \label{eq:velocitytilt}
\begin{aligned}
v^x_{\kv}&=v_{x}(\cos\theta+t),\\
v^y_{\kv}&=v_{y}\sin\theta.
\end{aligned}
\ee
The Jacobian determinant of transformation~\eqref{Polar2} is
$\mathcal{J}(\varepsilon,\theta)=\varepsilon/[\hbar^2 v_x v_y(1+t\cos\theta)^2]$, and for the DOS of a tilted Dirac system we find
\be\label{eq:dos2}
\begin{split}
\rho(\varepsilon)&=g \int \frac{d^2k}{(2\pi)^2}\delta(\varepsilon-\varepsilon_{\kv})\\
&=\frac{g \varepsilon}{2\pi \hbar^2 v_x v_y (1-t^2)^{3/2}},
\end{split}
\ee 
where $g$ is the band degeneracy factor.
\subsubsection{Vector mean free path  }
Here, again, we consider a random distribution of short-range impurity scatterers $V(r)=V_0\sum_j\delta(\rv-{\bf R}_j)$.
The zeroth-order (in the magnetic field) mean free paths for the conduction band are obtained as (see Appendix~\ref{app:II} for details)
\be \label{eq:tautilt}
\begin{aligned}
\lambda^{x,(0)}_{\kv}&=l_0^x  \left(1-t^{2}\right)^{3/2} \left(\frac{\cos\theta}{1- t \cos\theta}+\frac{t}{2}\right),\\
\lambda^{y,(0)}_{\kv}&= l_0^y\frac{(1-t^{2})^{3/2}}{1+t^{2}}\frac{\sin\theta}{1- t \cos{\theta}},
\end{aligned}
\ee
where $l^i_{0}= v_i\tau_0$, with $\tau_0= 4\pi\hbar^2v_{x}v_{y}/(u_{0}\varepsilon)$ ($u_{0}= \pi n_{imp}V_{0}^{2}/\hbar$).
With a finite magnetic field, we find the first-order correction to the mean free paths as
\be \label{eq:lxyt}
\begin{aligned}
\lambda^{x,(1)}_{\kv}&=l_{0}^{x}S_0(\varepsilon)(1-t^{2})^{3}\frac{\sin\theta}{2(1-t\cos\theta)} \\
&\quad \times \left[\frac{1}{1+t^{2}}+\left(\frac{1+t\cos\theta}{1-t\cos\theta}\right)^2\right],\\
\lambda^{y,(1)}_{\kv}&=-l_{0}^{y}S_0(\varepsilon)\frac{(1-t^{2})^{3}}{1+t^{2}}\frac{1}{2(1-t\cos\theta)}\\
&\quad \times \left[
\cos\theta+(\cos\theta-t)\left(\frac{1+t\cos\theta}{1-t\cos\theta}\right)^2\right],
\end{aligned}
\ee
with $S_0(\varepsilon)= e\tau_0 v_x v_y/ \varepsilon$. 
Eqs.~\eqref{eq:tautilt} and~\eqref{eq:lxyt} indeed satisfy the particle number conservation condition, i.e., Eq.~\eqref{eq:number} too.
Note that in the un-tilted (i.e., $t=0$) limit, these expressions simplify to
$\lambdav^{(0)}_{\kv}=\tau_0(v_x\cos\theta,v_y\sin\theta)$, and $\lambdav^{(1)}_{\kv}=\tau_0 S_0(v_x\sin\theta,-v_y\cos\theta)$, respectively.

\begin{figure}
\begin{tabular}{cc}
\hspace{-0.07\linewidth}\includegraphics[width=0.8\linewidth]{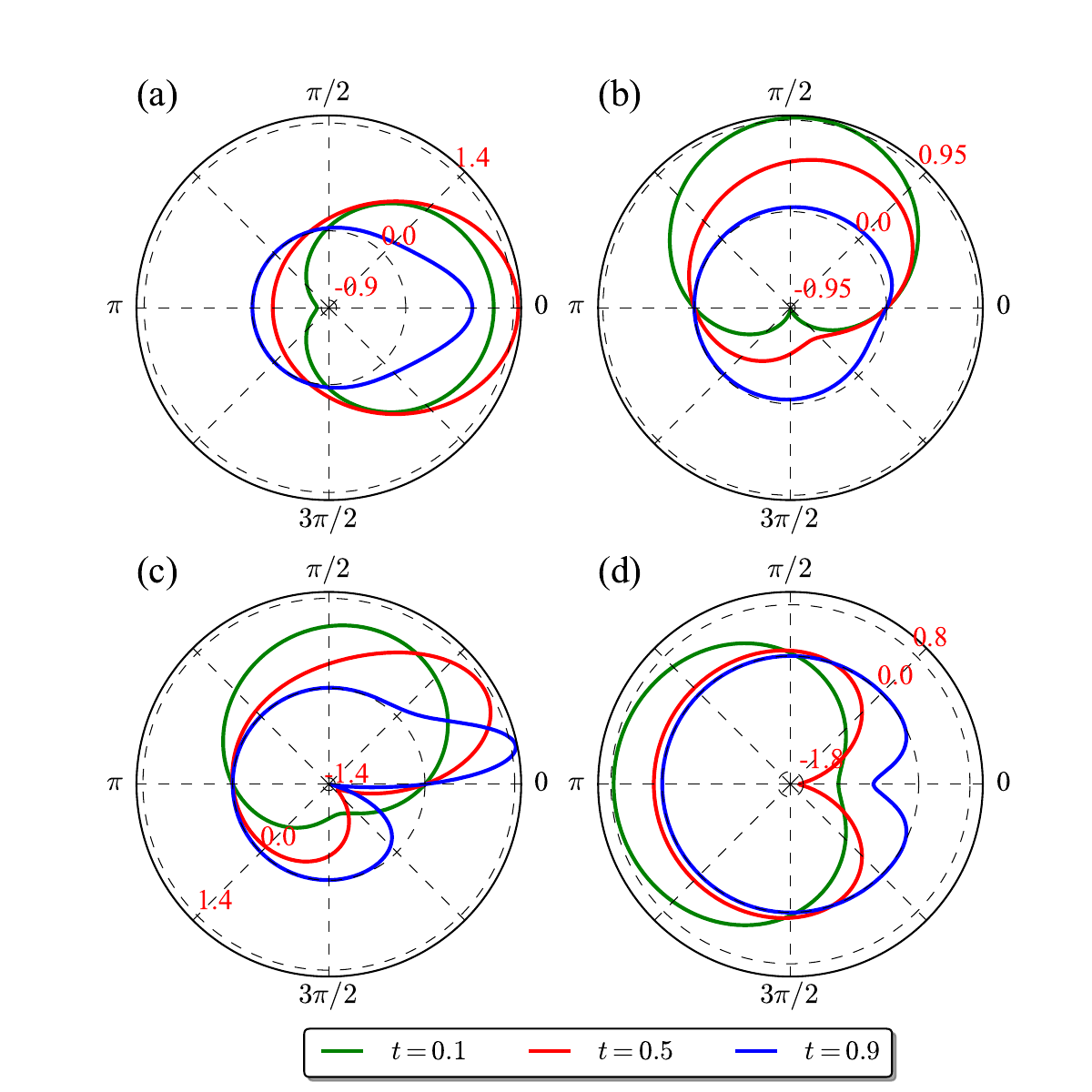}&
\hspace{-0.07\linewidth}\raisebox{0.2\linewidth}{\includegraphics[width=0.23\linewidth]{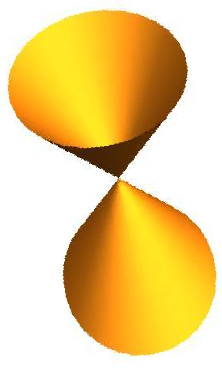}}
\end{tabular}
\caption{The polar plots of the mean free path of a two-dimensional system of tilted massless Dirac fermions along different directions and for varying tilt parameters. 
Top panels: zeroth order components $\lambda^{x,(0)}/l_0^x$ (a), and $\lambda^{y,(0)}/l_0^y$ (b). 
Bottom panels: first-order components
$\lambda^{x,(1)}/(l_0^xS_0)$ (c), and $\lambda^{y,(1)}/(l_0^y S_0)$  (d). 
The plot on the right is a schematic band structure of tilted massless Dirac fermions.
\label{fig:lambda_tilted}} 
\end{figure}

In Fig.~\ref{fig:lambda_tilted}, we illustrate the polar plots of the VMFP of a tilted system of massless Dirac fermions for varying tilt parameters. 

\subsubsection{Thermoelectric coefficients}
Now, it is straightforward to obtain different thermoelectric responses for tilted massless Dirac fermions.
For the electrical conductivity, from Eq.~\eqref{eq:general_transport} we find
\be \label{eq:sigmaiitilt}
\sigma =\sigma_0\frac{1-t^2}{1+t^2}
\begin{pmatrix}
(1+t^2) v_x   /v_y & -S_0(\mu) B \sqrt{1-t^2}\\
S_0(\mu) B \sqrt{1-t^2} &  v_y/v_x
\end{pmatrix},
\ee
where $\sigma_0= g e^2 v_x v_y/u_0$.
With the help of the thermoelectric conductivity 
\be \label{eq:alphatilt}
\alpha =-\dfrac{2\pi^2 k^2_B T }{3e \mu} \sigma_0S_0(\mu) B \frac{(1-t^2)^{3/2}}{1+t^2}
\begin{pmatrix}
 0 & 1\\
 -1 & 0 
\end{pmatrix},
\ee
we find the generalized Seebeck tensor 
\be
\label{eq:Stilt}
S=-\dfrac{2\pi^2 k^2_B T}{3e \mu} S_0(\mu) B\sqrt{1-t^2}
 \begin{pmatrix}
0 & v_y/(v_x(1+t^2))  \\
-v_x/v_y  & 0 \\
\end{pmatrix}.
\ee
It is interesting to observe that the Seebeck coefficients are zero along both $x$ and $y$ directions, as expected from the linear band dispersion. 
At the same time, the system displays a finite Nernst response, linear in the magnetic field, and a non-monotonically decreasing function of the tilt parameter $t$. 
The electrical conductivity and Nernst responses vanish for $t=1$ in our model.

In TABLE~\ref{table:tilted}, we summarize the energy dependence of different magneto-thermoelectric responses of a tilted two-dimensional massless Dirac system. The results obtained from the CRTA are also given in the table.
Again, CRTA's performance is very poor, calling for theories that take into account the energy and directional dependence of the mean-free paths.

\begin{table}
\caption{The chemical potential dependence of charge conductivity $\sigma_{ij}$ and Seebeck $S_{ii}$, and Nernst $S_{i\neq j}$ coefficients, of a tilted two-dimensional massless Dirac fermions obtained from the vector mean free paths (VMFP) method. The constant relaxation time approximation (CRTA) results are also provided for comparison. \label{table:tilted}}
\begin{tabular}{|c|c|c|c|c|c|c|c|c|}
\hline
Approach& $\sigma_{xx}$ & $\sigma_{yy}$  & $\sigma_{xy}(-\sigma_{yx})$   & $S_{xx}$ & $S_{yy}$  & $S_{xy}$  & $S_{yx}$\\
\hline
VMFP & $\mu^0$ & $\mu^0$ &$\mu^{-2}$ & 0 & 0 & $\mu^{-3}$ & $\mu^{-3}$\\ 
\hline
CRTA & $\mu$ & $\mu$ & $\mu^0$ & $\mu^{-1}$ & $\mu^{-1}$ & $\mu^{-2}$ & $\mu^{-2}$\\
 \hline
\end{tabular}
\end{table}

\section{Summary and Conclusions}\label{sec:conclusion}
In summary, employing the vector mean free path approach, we investigated the electronic contribution to the magneto-thermoelectric properties of anisotropic two-dimensional materials within the semiclassical Boltzmann formalism.
This approach provides an exact Ansatz for the non-equilibrium distribution function of the carriers to linear order in the external electric field and temperature gradient. In theory, it can be solved recursively to the desired order in the strength of the external magnetic field. 

For two representative two-dimensional anisotropic semimetals, i.e., semi-Dirac semimetal and tilted massless Dirac fermions, considering short-ranged impurity scatterings, we obtained the low-temperature linear magneto-thermoelectric responses as functions of the Fermi energy and to leading order in the magnetic field.
We also compared our results with those obtained within the widely used constant relaxation time approximation for the Boltzmann transport equation. We find significant qualitative differences for all components of thermo-magnetic responses in both of our studied models. This suggests that the appropriate incorporation of the energy and direction dependence of the non-equilibrium distribution function is crucial for correctly predicting the magneto-thermoelectric properties of anisotropic materials.

It would be interesting to investigate the effects of anomalous velocity in systems with non-zero Berry curvature~\cite{Suh_NJP2022} on the magneto-thermoelectric properties. The effects of other scattering mechanisms, particularly the inelastic scattering of electrons from phonons, require further inspection.

\acknowledgements
SHA thanks Reza Asgari, Ali G. Moghaddam, and Jahanfar Abouie for helpful discussions.
This work is supported by the research council of the Institute for Advanced Studies in Basics Science (IASBS), Zanjan, Iran.

\appendix
\section{Calculation of the  vector mean free path for semi-Dirac  semimetal}\label{app: free path}\label{app:I}
In this Appendix, we provide the details of obtaining the components of the VMFP $\lambda^{i,(s)}_{\kv}$ for a 2D semi-Dirac semimetal.

As the first step, we start with the zeroth order components $\lambda^{i,(0)}_{\kv}$, that is related to the zero magnetic field relaxation times $\tau^i_{\kv}$, as  $\lambda^{i,(0)}_{\kv}=v^i_{\kv}\tau^i_{\kv}$.  
The transition rate between the two states of the semi-Dirac system is
\be \label{eq:rate}
w_{\kv ,\kv'}=u_0 \left[1+|\cos\theta ||\cos\theta' |+\sin\theta \sin\theta' \right]\delta(\varepsilon - \varepsilon'),
\ee 
and we can also easily find ${\overline w}_\kv=\sqrt{\tilde{\varepsilon}}\,[I_1 + I_2|\cos\theta|]/(2\tau_0)$, where
\be\label{eq:I}
\begin{aligned}
I_1&=\int_0^{2\pi}\dfrac{ \mathrm{d}\theta}{\sqrt{| \cos\theta |}}=4\sqrt{2}K(1/2)\approx 10.48823,\\
I_2&=\int_0^{2\pi} \mathrm{d}\theta\sqrt{| \cos\theta |}=4\sqrt{\frac{2}{\pi}}\,\Gamma^2(3/4)\approx 4.79256,
\end{aligned}
\ee
with $K(x)$ and $\Gamma(x)$ the complete elliptic integral of the first kind and the Gamma function, respectively.

Putting the expression for $w_{\kv ,\kv'}$ into Eq.~\eqref{eq:relaxtimeaniso1}, we surmise
\be \label{eq:tau_semidirac}
\begin{aligned}
\widetilde{\tau}^x_\kv\left(I_1+I_2|\cos\theta |\right)
&=1+\frac{a^x_0+a^x_1 |\cos\theta |+a^x_2\sin\theta}{\cos\theta\sqrt{| \cos\theta |}},\\
\widetilde{\tau}^y_\kv\left(I_1+I_2|\cos\theta |\right)
&=1+\frac{a^y_0+a^y_1\cos\theta + a^y_2 \sin\theta}{\sin\theta},
\end{aligned}
\ee
where $\widetilde{\tau}^{i}_\kv=\sqrt{{\widetilde \varepsilon}}\tau^{i}_\kv/\tau_0$ (${\widetilde \varepsilon}$ and $\tau_0$ for semi-Dirac semimetal are introduced in the subsection~\ref{sec:responsesemi}). 

Upon the substitution of the expressions for the relaxation times from Eq.~\eqref{eq:tau_semidirac}  into Eq.~\eqref{eq:relaxtimeaniso1}, we find
\be\label{eq:as}
\begin{split}
a^i_0=0; \quad
a^i_1=0; \quad
a^x_2=0; \quad
a^y_2=\dfrac{I_3}{1-I_3},
\end{split}
\ee
where 
\be
I_3=\int_0^{2\pi} \mathrm{d}\theta\dfrac{\sin^2\theta}{\sqrt{| \cos\theta |}(I_1+I_2| \cos\theta |)}\approx 0.59862 .
\ee
Now, replacing the coefficients $a^i_n$ from Eq.~\eqref{eq:as} into Eq.~\eqref{eq:tau_semidirac}, we find the final expressions for the relaxation times along the $x$ and $y$ direction as
\be \label{eq:tausemi}
\begin{aligned}
\tau^x(\varepsilon,\theta)
&=\frac{ \tau_0}{\sqrt{\widetilde{\varepsilon}}\left(I_1+I_2|\cos\theta |\right)},\\
\tau^y(\varepsilon,\theta)
&=\frac{\tau_0}{\sqrt{\widetilde{\varepsilon}}(1-I_3)\left(I_1+I_2|\cos\theta |\right)}.
\end{aligned}
\ee
Now that we have $\lambda^{i,(0)}_\kv=v^i_\kv \tau^i_\kv$, we can solve
\be \label{eq:lams1}
\overline{w}_{\kv}\lambda_{\kv}^{i,(1)}=\sum_{\kv'}w_{\kv,\kv'}\lambda_{\kv'}^{i,(1)}-\frac{e}{\hbar}\hat{z} \cdot (\vv_{\kv}\times \nablav_\kv\lambda_{\kv}^{i,(0)}),
\ee 
to find the first-order corrections to the VMFP at the finite magnetic field. 
Similarly to what we did for the first-order components, after performing the summation over $\kv'$, we find
\be \label{eq9}
\begin{aligned}
\lambda_{\kv}^{i,(1)}&\left(I_1+I_2 |\cos\theta | \right)=
b^i_0+b^i_1 |\cos\theta |+b^i_2 \sin\theta \\
&-\dfrac{2e\tau_0}{\hbar \sqrt{\widetilde{\varepsilon}}}\left(v^x_\kv \partial_{k_y}\lambda_{\kv}^{i,(0)}-v^y_\kv \partial_{k_x}\lambda_{\kv}^{i,(0)}\right) .
\end{aligned}
\ee
We can find  the coefficients $b^i_0$, $b_1^i$, and $b^i_2$, through the Fourier expansion of   $\lambda_{\kv}^{i,(1)}$, that leads to
\be
b^x_2=\left(\dfrac{e\tau^2_0 v}{m \widetilde{\varepsilon}}\right)\dfrac{I_4}{1-I_3},
\ee
where 
\be \label{eq}
\begin{split}
I_4 &=\int_0^{2\pi} \mathrm{d}\theta \sqrt{|\cos\theta|}\sin^2\theta\dfrac{ 3I_1+I_2 |\cos\theta |}{\big(\, I_1+I_2 |\cos\theta |\, \big)^3}\\
&\approx 0.03101,
\end{split}
\ee
and all other coefficients are zero. This gives us Eq.~\eqref{eq:lxs} for $\lambda_{k}^{i,(1)}$.

Finally, in Eq. \eqref{sigmasd} for the electrical conductivity of semi-Dirac semimetals, we have introduced the following dimensionless coefficients 
\be\label{eq:ci}
c_{1}=\int_{0}^{2\pi}\mathrm{d}\theta \frac{\cos^{2}\theta \sqrt{|\cos\theta|}}{I_{1}+I_{2}\left|\cos\theta\right|}\approx 0.19682,
\ee
 $c_{2}=a_2^y=I_{3}/(1-I_{3})\approx 1.49144$, and $c_3=I_4/(2(1-I_3))\approx 0.03863$.

\section{Calculation of the vector mean free path for tilted massless Dirac fermions}\label{app:II}
This Appendix is devoted to the details of obtaining the vector mean free paths $\lambda^{i,(s)}_{\kv}$ for a 2D system of tilted massless Dirac fermions.
The transition rate between two eigenstates of the system in the presence of short-range impurities is 
 \be \label{}
 w_{\kv,\kv'}=u_0 \left[1+\cos\theta \cos\theta' +\sin\theta \sin\theta' \right]\delta(\varepsilon - \varepsilon^{'}),
 \ee
and we readily obtain $\overline w_{\kv}=2(1-t\cos\theta)/(\tau_0(1-t^2)^{3/2})$, with $\tau_0$ as defined in the subsection~\ref{sec:responsetilt}.

Upon the substitution of this expression into Eq.~\eqref{eq:relaxtimeaniso1}, and using the group velocities from Eq.~\eqref{eq:velocitytilt}, we find the relaxation times for a tilted Dirac semimetal
 \be \label{eq:relaxationtimetilit}
 \tau^i_\kv=\frac{(1-t^{2})^{3/2}}{1-t\cos\theta}\left[\dfrac{\tau_0}{2}+\frac{a^i_{0}+ a_{1}^{i}\sin\theta+a_{2}^{i}\cos\theta}{\upsilon^{i}_{\kv}}\right] .
 \ee
Next, by putting back this expression for $\tau^i_\kv$ into  Eq.~\eqref{eq:relaxtimeaniso1}, we find the coefficients $a^i_n$ as
\be
a^x_2=\frac{1}{2}\tau_0\upsilon_{x}(1-t^2); \quad
a^y_1=\frac{1}{2}\tau_0\upsilon_{y} \left(\frac{1-t^2}{1+t^2}\right),
\ee
and all other coefficients are zero. 
This gives the  relaxation times 
\be \label{eq:tautilt1}
\begin{aligned}
\tau^{x}_{\kv}&=\tau_0 \frac{(1-t^{2})^{3/2}}{1-t\cos\theta}\frac{t+(2-t^{2})\cos\theta}{2(t+\cos\theta)},\\
\tau^{y}_{\kv}&=\tau_0 \frac{(1-t^{2})^{3/2}}{1-t\cos{\theta}}\frac{1}{1+t^{2}}.
\end{aligned}
\ee
For the first-order corrections to the vector mean free paths, we use Eq.~\eqref{eq:lams1} and follow the same steps explained in the Appendix. \ref{app:I}, to find Eq.~\eqref{eq:lxyt} after some straightforward calculations.

\section{Magneto-thermoelectric responses of anisotropic materials within the constant relaxation time approximation}\label{app:CRTA}
In this Appendix, we take a constant (i.e., energy and direction-independent) relaxation time to approximate the non-equilibrium distribution function. Then, we show how such a crude approximation performs in studying the magneto-thermoelectric properties of our anisotropic model systems. 

Treating the collision term in the Boltzmann equation~\eqref{eq:BTE_lin} in the CRTA, we have
\be \label{eq:BTE_linRTA}
\vv_{\kv}\cdot {\bf F} =-\frac{g_\kv}{\tau_0}- \widehat \Omega_\kv g_\kv ,
\ee
where $\tau_0$ is the constant relaxation time (note its different definition from the main text). Substituting the general Ansatz in Eq. \eqref{eq:gk} for the non-equilibrium part of the distribution function into Eq. \eqref{eq:BTE_linRTA}, we find
\be \label{eq:BTE_linRTA1}
\tau_0 v^i_\kv=\Lambda^i_\kv+\tau_0\widehat\Omega_\kv\Lambda^i_\kv.
\ee
Now, similar to the main text, if we expand $\Lambda^i_\kv$ in the powers of the external magnetic field and equate the coefficients of the corresponding powers, we obtain
\be \label{eq:lams1C}
\begin{split}
\lambda_{\kv}^{i,(0)}&=\tau_0\upsilon_{\kv}^{i},\\
\lambda_{\kv}^{i,(s)}&=-\frac{e\tau_0}{\hbar}\hat{z} \cdot (\vv_\kv \times \nablav_\kv\lambda_{\kv}^{i,(s-1)}),\quad s\geq 1. 
\end{split}
\ee

\subsubsection{Two-dimensional semi-Dirac semimetal}
For semi-Dirac system, we substitute the group velocities from Eq.~\eqref{eq:velocity} into Eq.~\eqref{eq:BTE_linRTA1} and find
\be \label{eq:semidirac_lambda0RTA}
\begin{aligned}
\lambda^{x,(0)}_{\kv}&=l_0\cos\theta \sqrt{\widetilde{\varepsilon}\left|\cos\theta\right|},\\
\lambda^{y,(0)}_{\kv}&=l_0\sin\theta,
\end{aligned}
\ee
and
\be \label{eq:lxsRTA}
\begin{aligned}
\lambda^{x,(1)}_{\kv}=&\frac{3}{2}l_0 S_0\left|\cos\theta\right|\sin\theta,\\
\lambda^{y,(1)}_{\kv}=&-l_0 S_0\cos\theta\sqrt{{\left|\cos\theta\right|}/{\widetilde{\varepsilon}}},
\end{aligned}
\ee
for the zero and first-order mean-free paths.
Here, $l_0$, $S_0$, and $\widetilde{\varepsilon}$ are defined the same way as the main text (see subsection~\ref{sec:responsesemi}), albeit in terms of the constant phenomenological $\tau_0$. 
Using Eqs.~\eqref{eq:semidirac_lambda0RTA} and \eqref{eq:lxsRTA}, we find the electric conductivity
\be \label{sigmasdRTA}
\sigma=\sigma_0\begin{pmatrix}
 r_1 {\widetilde \mu}^{3/2} & - r_1 S_0 B{\sqrt{\widetilde \mu}}  \\
r_1 S_o B{\sqrt{\widetilde \mu}} & r_2{\sqrt{\widetilde \mu}} \\
\end{pmatrix},
\ee
where $\sigma_0=(ge^2\tau_0\varepsilon_0)/(2\pi\hbar)^2$, and the dimensionless constants $r_i$ are defined as
\be\label{eq:ri}
\begin{aligned}
r_{1}&=\int_{0}^{2\pi}d\theta \cos^{2}\theta \sqrt{|\cos\theta|}=\dfrac{3\Gamma^2(-1/4)}{10\sqrt{2\pi}}\approx 2.87554,\\
r_{2}&=\int_{0}^{2\pi}d\theta \dfrac{\sin^{2}\theta}{\sqrt{|\cos\theta|}}=\dfrac{8\sqrt{2}K(1/2)}{3}\approx 6.99215.
\end{aligned}
\ee
Now,  it is straightforward to find the generalized Seebeck tensor as
\be
\label{eq:semidiracRTA}
S=-\dfrac{\pi^2 k^2_B T}{6e \mu}
 \begin{pmatrix}
3  & 0 \\
-r_1 S_0 B /r_2  &1\\
\end{pmatrix}.
\ee
Comparing Eqs.~\eqref{sigmasdRTA} and~\eqref{eq:semidiracRTA} with Eqs.~\eqref{sigmasd} and~\eqref{eq:semidirac}, respectively, we see that the CRTA fails even to provide qualitatively reasonable results. 
In TABLE~\ref{table:semiDirac}, we have summarized the chemical potential dependence of different magneto-thermoelectric responses obtained from CRTA and what we found from the VMFP approach.

\subsubsection{Two-dimensional system of tilted massless Dirac fermions}
For a tilted Dirac system, we take the group velocities from Eq.~\eqref{eq:velocitytilt} and find
\be \label{eq:tautiltRTA}
\begin{aligned}
\lambda^{x,(0)}_{\kv}&=l_0^x  \left(t+\cos\theta\right),\\
\lambda^{y,(0)}_{\kv}&=l_0^y\sin\theta,
\end{aligned}
\ee
and
\be \label{eq:lxytRTA}
\begin{aligned}
\lambda^{x,(1)}_{\kv}&= l_{0}^{x}S_0\left(1+t\cos\theta\right)^2\sin\theta,\\
\lambda^{y,(1)}_{\kv}&=- l_{0}^{y}S_0\left(1+t\cos\theta\right)^2\cos\theta,
\end{aligned}
\ee
for the zero and first-order mean free paths, respectively. Here, $l_0^i$ and $S_0$ are as defined in subsection~\ref{sec:responsetilt}, but in terms of the constant relaxation time $\tau_0$,

For the electrical conductivity, we obtain
\be \label{eq:sigmaiitiltRTA}
\sigma =\sigma_0
\begin{pmatrix}
\dfrac{2(1-\sqrt{1-t^2})v_x}{t^2v_y} & -S_0(\mu) B\\
S_0(\mu) B & \dfrac{2(1/\sqrt{1-t^2}-1)v_y}{t^2v_x}
\end{pmatrix},
\ee
with $\sigma_0=(ge^2\tau_0\varepsilon)/(4\pi\hbar^2)$. 
At the end, we find the Seebeck tensor as
\be
\label{eq:StiltRTA}
S=-\dfrac{\pi^2 k^2_B T}{6e \mu}
 \begin{pmatrix}
2 & \dfrac{S_0(\mu) B t^2v_y}{(1-\sqrt{1-t^2})v_x}  \\
-\dfrac{S_0(\mu) B t^2v_x}{(1/\sqrt{1-t^2}-1)v_y}  & 2 \\
\end{pmatrix}.
\ee
Again, the results are in significant contrast with our VMFP results. The summary of corresponding results is given in TABLE~\ref{table:tilted}.


\end{document}